\documentclass[11.5pt preprint]{aastex}

\usepackage{lscape}
\usepackage{graphicx}
\usepackage{dcolumn}
\usepackage{bm}
\usepackage{natbib}

\begin{document}


\title{The Evolution of Density Structure of Starless and Protostellar Cores}  

\author{Chao-Ling Hung \& Shih-Ping Lai}
\affil{Institute of Astronomy and Department of Physics, National Tsing-Hua University, Hsinchu 30013, Taiwan}
\email{lynn0701@gmail.com, slai@phys.nthu.edu.tw}
\and
\author{Chi-Hung Yan}
\affil{Institute of Astronomy and Astrophysics, Academia Sinica, Taipei 10617, Taiwan \& \\ National Taiwan Normal University, Taipei 10617, Taiwan}
\email{chyan@asiaa.sinica.edu.tw}


\begin{abstract}
We present a near-infrared extinction study of nine dense cores at evolutionary stages between starless to Class I.   Our results show that the density structure of all but one observed cores can be modeled with a single power law $\rho \propto r^p$ between $\sim$ $0.2R-R$ of the cores.  The starless cores in our sample show two different types of density structures, one follows $p\sim -1.0$ and the other follows $p\sim -2.5$, while the protostellar cores all have $p\sim -2.5$.  The similarity between the prestellar cores with $p\sim -2.5$ and protostellar cores implies that those prestellar cores could be evolving towards the protostellar stage.   The slope of  $p\sim -2.5$ is steeper than that of an singular isothermal sphere, which may be interpreted with the evolutionary model of cores with finite mass. 


\end{abstract}


\section{\label{sec:level1}Introduction\protect\\}
To understand the nature of star formation processes, it is crucial to measure the fundamental physical parameters of molecular cores during their evolution. Density structure is one of the most important parameters which can provide critical information to test theoretical models.
The simplest model for a pre-collapsing core is a Bonnor-Ebert sphere (hereafter BES; Bonnor 1956 and Ebert 1955), which describes an isothermal, non-magnetized, non-rotating, hydrostatic core bounded by external pressure. 
For a collapsing core, the standard model is inside-out collapse model which  predicts the density $\rho \propto r^{-2}$ before the gravitational collapse sets in and  $\rho \propto r^{-1.5}$ in the inner free fall region (Shu 1977). 
Vorobyov and Basu (2005) find the power-law index of density profile becomes much steeper than $\rho \propto r^{-2}$ in the outer region of the core if the core has finite mass reservoir. 

Two methods have been commonly used to map the density structure of cores. One is to measure the thermal dust emission at millimeter or submillimeter wavelengths using single dishes or interferometers, and the other is to measure the extinction of the background stars at near infrared (NIR) wavelengths.  Large surveys of core density structure can be conducted relatively easily using single dishes, but the resolutions are often limited and the density profile can depend on observing wavelength (Shirley et al. 2000; Kauffmann et al. 2008).  Interferometers can be used to achieve high resolution in the inner-most region, but the overall structure are often resolved out (Harvey et al. 2003a).  Given enough background stars and integration time, NIR extinction observations have the best potential to achieve both high resolution and large scale mapping of the core density structure (Alves et al. 2001).  

Previous studies have shown that BES or power-law density profiles are good descriptions for density profiles of molecular cores.  
Kandori et al.\ (2005) found that Bok Globlules can be described as either gravitationally stable or unstable BESs with NIR extinction observations.  Shirley et al.\ (2000) found that the density profiles of 21 low-mass cores from starless to Class I stages can be fitted with power laws $\rho \propto r^{-1.5}$ to $\rho \propto r^{-2.6}$ in the resolved regions with assumed dust temperature $T_d(r) \propto r^{-0.4}$.   
It is unclear, however, whether or not the variations in the observed density structure are due to evolutionary effects. 
In this study, we focus on investigating the evolutionary trend of core density profiles with NIR extinction observations. The observations and data reduction processes are described in \S 2. The derived visual extinction ($A_V$) distribution and the radial density profiles are presented in \S 3. We discuss the implications from our results in \S 4.

\section{\label{sec:level1}Observation and Data Reduction\protect\\}
We observed five isolated cores (CB68, L483, CB188, L158 and B59) and two fields in Serpens Molecular Cloud in the H and Ks bands using the Wide-field InfraRed Camera (WIRCam) on Canada-France-Hawaii Telescope (CFHT) in 2008 April and May. 
These fields are located in low Galactic latitude and hence they have abundant background stars.
The isolated cores and the selected dense cores in two Serpens fields have relatively round shape and contains only a protostar (Class 0 or I), except for B59. B59 is a cluster forming core and is chosen for the comparison between single star and cluster formation. 

For each observing field, we performed in total a 17.5 minute integration (5 sets of 21 dithered images) in H band and a 3.5 minute integration (one set of 21 dithered images) in Ks band. 
Standard bias subtraction, flat-fielding, bad pixel masking, and sky background subtraction were performed using `I`iwi system, the WIRCam pipeline \footnote{
http://www.cfht.hawaii.edu/Instruments/Imaging/WIRCam/IiwiVersion1Doc.html}.    
The sky background of each image was subtracted based on the co-moving averaged frame with length of 15 minutes. 
We combined all images of each observing field and used SExtractor package (Bertin and Arnouts 1996) for the source extraction and photometry except for the extremly crowded CB188 field for which we used DAOPHOT package (Stetson 1987). 
The photometry results from both methods were calibrated using the point sources with 2MASS detections between 13 to 15 magnitude. The typical limiting magnitudes with 3 $\sigma$ detection are about 20.7 and 19.5 in H and Ks band.  

\section{\label{sec:level1}Analysis and Results\protect\\}
\subsection{\label{sec:level2}The Visual Extinction $A_V$}
The $A_V$ value is determined from near-infrared color excess $E_{H-Ks}$ of stars in the H and Ks bands (NICE method, Lada et al. 1994), i.e., $A_V=r^{H,Ks}_{v}E_{H-Ks}$.
$r^{H-Ks}_{v}$ is given by a dust reddening law and $E_{H-Ks}$ is evaluated from $(H-Ks)_{observed}-(H-Ks)_{intrinsic}$. 
We use $r^{H-Ks}_{v}=19.245$ as derived from the Rv=5.5 model in Weingartner and Draine (2001, WD). Recent studies suggest that the reddening law in dense cores is better described with WD's Rv=5.5 model rather than their Rv=3.1 model which fits the reddening law derived from more diffused regions (Indebetouw et al. 2005; Chapman et al. 2009). $E_{H-Ks}$ can be obtained from observations once $(H-Ks)_{intrinsic}$ of the star is known.
We adopt $(H-Ks)_{intrinsic}=0.13$ mag (Lada et al. 1994) and assume it is a constant for all observed fields. 

The detected point sources in our fields are not necessary stars; they could be young stellar objects or distant galaxies. 
We cannot perform the standard $JHK_s$ color-color diagram to select stars since we did not observe our target sources in J band to save observation time. 
We can however take advantage of the existing catalogs from the Spitzer c2d Legacy Project (Evans et al. 2003) to select bona-fide stars detected in our fields. 
The c2d Project has developed a sophisticated source classification procedure based on the 2MASS and Spitzer data (Harvey et al. 2007).  
If the point sources are not contained in c2d catalog, we assume them to be stars.
The derived extinction maps are shown in Figure 1.

\subsection{\label{sec:level2}The Density Profile}
We assume the three dimensional shape of our observing cores are spheres and derive their density profiles. We first determine the center, radius, and background extinction of the cores (Table 1), then annularly average the derived extinction to produce the extinction profile (Figure 2).  

The centers of the four cores with each a single embedded star (CB68, L483, CB188 and Serp-Bolo20) are set to be the position of their associated protostar since the protostar and the peak of $A_V$ distribution almost coincide. The center of the four starless cores (L158, Serp-Bolo1, Serp-Bolo5 and Serp-Bolo19) and the cluster forming core, B59, is set at the peak of their $A_V$ distribution. 
We set the annular averaging step as $\Delta r$ and ensure that each step contains at least five detections to obtain properly the average $A_V$.
Foreground star contamination may affect the derived extinction. We expect, however, the foreground star fraction is small since all the selected regions are nearby ($\sim 200$ pc).
If foreground stars contaminate the data in $\Delta r$, the derived $A_V$ of foreground stars will deviate from the median of the data. Therefore,
to remove the possible bias from foreground stars,
we compute the median ($m_{A_V}$) and the standard deviation ($\sigma_{A_V}$) of $A_V$ distribution in each annulus and remove the sources with $A_V$ larger than $m_{A_V}+1.5$ $\sigma_{A_V}$ or smaller than $m_{A_V}-1.5$ $\sigma_{A_V} $.
Then we annularly average $A_V$ in each $\Delta r$.

At a certain distance from the center, the $A_V$ approaches nearly a constant value which is defined as the background $A_V$. We calculate the average background $A_V$ (BG) and the standard deviation (SD) in the background region.  The radius of the cores is defined as the distance from center to the position where the $A_V$ exceeds $BG+3SD$ and the $A_V$ profiles are obtained by subtracting BG. The relationship between the $A_V$ and $H_{2}$ column density are derived from $(N(H_I)+2N(H_2))/E_{B-V}=5.8\times10^{21}$ cm$^{-2}$ mag$^{-1}$ (Bohlin et al.\ 1978). Assuming all hydrogen in the dense cores is in $H_2$ form and adopting the $R_V$=5.5 reddening law (where $R_V\equiv A_V/E_{B-V}$), we obtain $N(H_2)/A_V=5.27\times10^{20}$ cm$^{-2}$ mag$^{-1}$.

\subsection{\label{sec:level3}Modeling}
The observed column density profiles can be modeled with various functions that represent different inherent physical conditions. 
Bonnor-Ebert spheres are commonly used for presumed hydrostatic cores with flat density profiles in the center and are truncated at radius $R$.  
On the other hand, the solutions from different collapsing models could be as simple as a single power-law function or as complex as non-analytic functions.
In general, these solutions can have approximate forms described by a power-law function with different power-law indices at different positions from the center.
In this work, we model the density structure of dense cores with a single power-law profile as the first approach and further examine the power-law index variation within the density profile in $\S 4.2$. 

Assuming the volume density $\rho(r)\propto r^{p}$, the observed column density $N(w)$ at the projected radius $w$ in the plane of the sky is
\begin{equation}
N(w)=2\int_{w}^{R}{\rho(r)\frac{r}{\sqrt{r^2-w^2}}}dr.
\end{equation}
The best fit power-law index $p$ in $\rho(r)$ is obtained by minimizing the $\chi^2$ between derived density profile and power-law density model.
We adopt the Bootstrap test to determine the uncertainty in $p$ (see astronomical application in Wallin et al. 2007 and ``Numerical Recipes'' by Press et al. 1993).   
The main idea of the Bootstrap test is to find the uncertainty of any operation as the dispersion of the results from the same operation on a large number of simulated data sets.  We produce 100 sets of simulated density profiles for each core, where each data point in the simulated profiles is generated from a Gaussian random number generator with the mean and standard deviation of the probability distribution equal to the mean and the uncertainty of the observed data.  We therefore obtain the uncertainty in $p$ from performing our fitting procedure on each simulated data set and measuring the standard deviation of the fitting results for each core.
The best-fit results are listed in Table 1. Figure 2 shows the best-fit results overploted on the observed data.

Most of our samples have data in the range of $0.2R-R$ except for Serp-Bolo1 ($\sim 0.3R-R$) and CB188 ($\sim 0.06R-R$). 
To compare fairly the average power-law index of each core with the same fitting range, 
we further fit all samples in $0.3R-R$ and fit CB188 in $0.2R-R$ and $0.3R-R$.
Figure 3 shows the original best-fit results and the results obtained with the fitting range in $0.2R-R$ and $0.3R-R$.
The normalized $\chi$ square, $\chi^2_N$, of the fitting results are all smaller than one because the dispersion of the data in the same radius are in general larger than the fitting uncertainty. It is common to use the measurement uncertainty as the weighting in fitting processes in many works (e.g. Harvey et al. 2003b). We use however the dispersion of annularly averaged data as the error bar here to include the uncertainty in assuming spherical structure. This assumption reduces the fitting $\chi^2_N$, but increases the uncertainty in $p$.

\section{\label{sec:level1}Discussion\protect\\}
\subsection{\label{sec:level2}The evolution of the density structure}
We investigate the power-law index variation of the density structure
at different evolutionary stages. 
To increase the size of our sample, we include the results from
several similar infrared extinction studies, including L694-2 (Harvey
et al. 2003b), B335 (Harvey et al. 2001),
Thumbprint Nebula and DC303.8 (Kainulainen et al. 2007).
In these works, the
extinction values are all determined from H-K color excess and the
density profiles are all fitted with single power law, which is the
same as our approach.
Although the computer codes used to do the fitting may not be
identical, we expect that these differences will not be significant.

The six starless cores in this sample show two different types of density structure. Serp-Bolo19 and L694-2 have the best-fit power-law index $\sim$ -2.5 (hereafter ``steep type''), while other four cores (Serp-Bolo5, Serp-Bolo1, L158 and Thumbprint Nebula) have the power-law index $\sim$ -1.0 to -1.4 (hereafter ``shallow type''). 
This index range in shallow type reduces to $\sim$ -1.0 to -1.2 if we reduce the fitting range to  $0.2R - R$.
The difference between these two types is significant even after considering the uncertainty of best-fit results and it is obtained under fair comparison with similar fitting range.
On the other hand, the power-law indices of Class 0 samples do not have clear two different types as starless cores but a range about -1.9 to -2.6. 
The two samples in Class I stage show two different power-law indices again (-2.7 in Serp-Bolo20 and -1.2 in CB188).
This discrepancy may result from different fitting ranges. Indeed, their best-fit results become similar to Class 0 sources after considering only a fixed fitting range within $\sim 0.2R-R$ (Figure 3).

For the starless cores, the $A_V$ difference between the center and the edge also show a dichotomous nature. The shallow type starless cores have much smaller $A_V$ difference ($\sim$ 22 mag) compared to the steep type starless and protostellar cores ($\sim$ 45 mag). 
If all starless cores in our sample eventually evolve into protostellar cores, our results suggest that the starless cores with shallower density structure will evolve into steeper ones via certain core-forming processes and their density structure have a averaged $p\sim-2.5$ during the formation of the protostars. One way to examine if the starless cores will evolve into the protostellar cores is to probe their kinematic structure.  Lee et al. (2004)  shows that both L158 (shallow type) and L694-2 (steep type) have infall signatures in the molecular spectra, which are consistent with our suggestions that some shallow type starless cores will evolve to the steep type during the formation of protostars.

\subsection{\label{sec:level2}Power-law index variation within density profile}

Some core collapsing models predict a power-law index variation within the density profile. For example, a discontinuity in power-law indices of the density profile separates the interior infalling region from the exterior static region, and the discontinuity will propagate outward with sound speed during collapse (e.g. Shu 1977, Fatuzzo et al. 2004).    Since the crossing time of the discontinuity ($\approx 10^5$ yr, estimated with $T=10K$, $m=2.33$ $amu$ and $R=0.1$ $pc$) is comparable to the statistical lifetime of Class 0 and Class I stages ($\approx 10^4$ yr and $\approx 10^5$ yr, Andre et al. 2000),  we should be able to see the discontinuity in the observed density profiles in Class 0 or Class I stages. 

The best-fit results with fitting range of $0.3R-R$ showed in Figure 3 are in general steeper than fitting range of $0.2R-R$, suggesting the power-law index may vary within core density profile and specifically, it may become steeper near the edge. 
We examine this trend by modified fitting processes. 
Considering the simplest case, a core has a two power-law volume density profile with the discontinuity at $r=R'$. Its column density profile within projected radius $R'$ will be an integration of the volume density over regions with both power-law indices along the line of sight while the column density outside of $R'$ will only reflect the power-law index in the outer region.  
Therefore, we can obtain the best-fit power-law index $p_1$ in the region $R'<r<R$ and then use $p_1$ as a known parameter to determine the best-fit power-law index $p_2$ in the $r<R'$ region.  $R'$ is the radius chosen where the single power-law fitting in the region of $R' - R$ reaches minimum $\chi^2_N$.

The feasibility of the above procedures, however, is limited due to the large uncertainty of annularly averaged data. 
When we vary $R'$ from outer region toward inner region, we may not be able to find obvious minimum $\chi^2_N$ and the best fit $p_1$ may vary significantly.
If we set the criteria for variation of best-fit $p_1$ smaller than $\pm0.3$ (i.e., the typical uncertainty in best-fit single power-law profile), 
L483 is the only possible candidate that shows deviation from a single power-law profile.
In L483, the $\chi^2_N$ of the two power-law model are $\sim$ six times smaller than the $\chi^2_N$ of single power-law profile.
This difference, however, is not significant since the best fit results of two models are both obtained with $\chi^2_N<1$.
The best-fit two power-law profile of L483 has the steep power index $p_1\sim-3.9$ at $r>R'$ $(\sim0.5R)$ and $p_2\sim-1.7$ at $r<R$.
This result shows the possible power-law index variation in a Class 0 sample,
the power-law index $-1.7$ in inner region is consistent with the power-law index in the infall region of inside-out collapsing scenario. 
The power-law index, however, becomes steeper than $-2$ at $r>0.5R$ suggesting other effects (e.g. see the discussions in \S 4.3) besides inside-out collapse should be considered in outer parts of the core.

\subsection{\label{sec:level3}The implication for core evolution}
Our results show that the density profiles evolve from a shallower one with $p\sim-1.0$ to a steeper one with $p\sim-2.5$ in the epoch of core formation and the steepened density profile is sustained during the formation of protostar.
Furthermore, the procedures for investigating the power-law index variation within a core suggest a steep index may present in the outside part of the core.

One explanation for these results is that a core has to form a finite mass object before the inside-out collapse sets in.  Vorobyov and Basu (2005, VB05) considered the effect of a finite mass reservoir during the core-forming and collapse phases. They found that an equivalent ``rarefaction wave'' is generated at the core boundary and propagates inward, which rapidly reduces the infall rate and steepens the density profile to $p<-2$ (Figure 1 in VB05).  
Our results show a similar evolutionary trend with a steep power-law index seen in the outer region of our samples, as described in VB05. 
In addition, the VB05 model predicts that the infall velocity increases toward the center and then decreases when the velocity reaches a peak at $\sim -0.2$ to $-0.4$ km s$^{-1}$. A recent study on the velocity field of two starless cores show results consistent with the VB05 model 
(Figure 2 in Lee et al. 2007).

The physical reason for the ``finite mass effect'' could originate from the boundary of supercritical and subcritical region. For example, Tassis and Mouschovias (2007, TM07) show that during core formation via ambipolar diffusion, the magnetically supported mass cannot replenish the inner supercritical infall mass, resulting in a steep power-law index in the outer parts of the core (Figure 2d in TM07). 
Therefore, we may still see the cores smoothly blend in with the background but not confined by sharp edges as the literal meaning of finite mass.
The origin of the ``finite mass effect'' and how significant is it for the star formation needs further investigation via observations on the density structure and also velocity structure of the cores at various evolutionary stages.





\acknowledgments
We thank Shantanu Basu and Fred C. Adams for their insightful discussion on the first draft of this paper.
This work is based on observations obtained at the Canada-France-Hawaii Telescope (CFHT) which is operated by the National Research Council of Canada, the Institut National des Sciences de l'Univers of the Centre National de la Recherche Scientifique of France, and the University of Hawaii. Access to the CFHT was made possible by the Institute of Astronomy and
Astrophysics, Academia Sinica, Taiwan. CLH and SPL are supported by National Science Council of Taiwan under grant NSC 96-2112-M-007-019-MY2 and NSC 98-2112-M-007-007-MY3.

\clearpage
\begin{landscape}
\begin{table}
\footnotesize
\begin{center}
\caption{The Basic Information and Fitting Results}
\label{info}
\begin{tabular}{lllccccccccc}
\tableline\tableline
Name & 
\multicolumn{2}{c}{Center Position} & 
\multicolumn{1}{c}{Distance} & 
\multicolumn{1}{c}{Class} & 
\multicolumn{1}{c}{Peak, BG Av} & 
\multicolumn{1}{c}{$\Delta r$} & 
\multicolumn{1}{c}{$D_{in}, R$} & 
\multicolumn{1}{c}{Mass} & 
\multicolumn{1}{c}{Best Fit} & 
\multicolumn{1}{c}{$\chi^2_N$}&  
\multicolumn{1}{c}{Reference\tablenotemark{e}}  
\\
 & 
\multicolumn{1}{c}{R.A.($^{\rm h}$ $^{\rm m}$ $^{\rm s}$)} & 
\multicolumn{1}{c}{Dec.($^{\circ}$ $^{'}$ $^{''}$)} & 
\multicolumn{1}{c}{(pc)} & 
\multicolumn{1}{c}{} & 
\multicolumn{1}{c}{(mag,mag)} & 
\multicolumn{1}{c}{($''$) } & 
\multicolumn{1}{c}{(pc,pc)} & 
\multicolumn{1}{c}{$M_{\odot}$} & 
\multicolumn{1}{c}{$p$} & 
\multicolumn{1}{c}{}&  
\multicolumn{1}{c}{}  
\\
\tableline
L158&16 47 22.2 & $-$13 59 23&$125$&Starless&$33.8,10.6$&$10$&$0.02,0.11$&$2.20$&$-1.1\pm0.3$&$0.67$& \\

Serp-Bolo1&18 28 23.1 & $+$00 26 34&$230$&Starless\tablenotemark{a} &$37.2,14.7$&$3.5$&$0.01,0.03$&$0.40$&$-1.4\pm0.7$&$0.17$&  \\
 
 Serp-Bolo5&18 28 48.3 & $+$00 14 51&$230$&Starless\tablenotemark{a}&$40.2,17.0$&$5.5$&$0.01,0.06$&$0.84$&$-1.0\pm0.5$&$0.08$&  \\
 
 Serp-Bolo19&18 29 31.5 & $+$00 26 49&$230$&Starless\tablenotemark{a}&$56.7,12.6$&$6$&$0.03,0.14$&$6.65$&$-2.5\pm0.6$&$0.05$& \\
 
 CB68&16 57 29.6 & $-$16 09 23&$150$&Class 0\tablenotemark{b}&$41.9,5.1$&$10$&$0.02,0.11$&$4.33$&$-2.6\pm0.3$&$0.15$& \\
 
 L483&18 17 30.0 & $-$04 39 40&$200$&Class 0\tablenotemark{c}&$42.3,4.2$&$10$&$0.03,0.2$&$15.3$&$-2.5\pm0.2$&$0.2$& \\
 
 Serp-Bolo20&18 29 31.9 & $+$01 19 01&$230$&Class I\tablenotemark{a} &$36.2,13.2$&$6$&$0.03,0.13$&$7.99$&$-2.7\pm0.6$&$0.05$& \\
 
 CB188&19 20 15.8 & $+$11 35 53&$300$&Class I\tablenotemark{d}&$27.2,7.0$&$6$&$0.01,0.17$&$3.78$&$-1.2\pm0.1$&$0.43$& \\
 
 B59&17 11 23.8 & $-$27 26 03&$125$&Cluster&$>50,7.4$&$15$&$0.03,0.19$&$9.70$&$-2.3\pm0.2$&$0.20$& \\
 \tableline 
 \tableline 
  Thumbprint&12 44 55.6 & $-$78 48 16&$125$ &Starless&$-$&$-$&$-$&$-$&$-1.19\pm0.09$&$1.13$&A \\
  
   L694-2&19 41 04.4 & $+$10 57 01&$125$ &Starless&$-$&$-$&$-$&$-$&$-2.6\pm0.2$&$1.11$&B \\
   
  DC303.8-14.2&13 07 38.0 & $-$77 00 21&$125$ & Class 0&$-$&$-$&$-$&$-$&$-2.0\pm0.04$&$1.58$&A \\
  
 B335&19 37 00.9 & $+$07 34 10&$125$ &Class 0&$-$&$-$&$-$&$-$&$-1.9\pm0.07$&$3.37$&C \\

 \tableline
\end{tabular}
\tablenotetext{a}{Identified by Enoch et al. (2007)}
\tablenotetext{b}{Classified by Vall{\'e}e et al. (2007)}
\tablenotetext{c}{Classified by Tafalla et al. (2000)}
\tablenotetext{d}{Classified by Kauffmann et al. (2008)}
\tablenotetext{e}{Results from other simlar works: (A)Kainulainen et al. (2007);  (B)Harvey et al. (2003b); (C)Harvey et al. (2001)}
\end{center}
\end{table}
\end{landscape}

\begin{figure*}
	\centering
	\includegraphics[width=\textwidth]{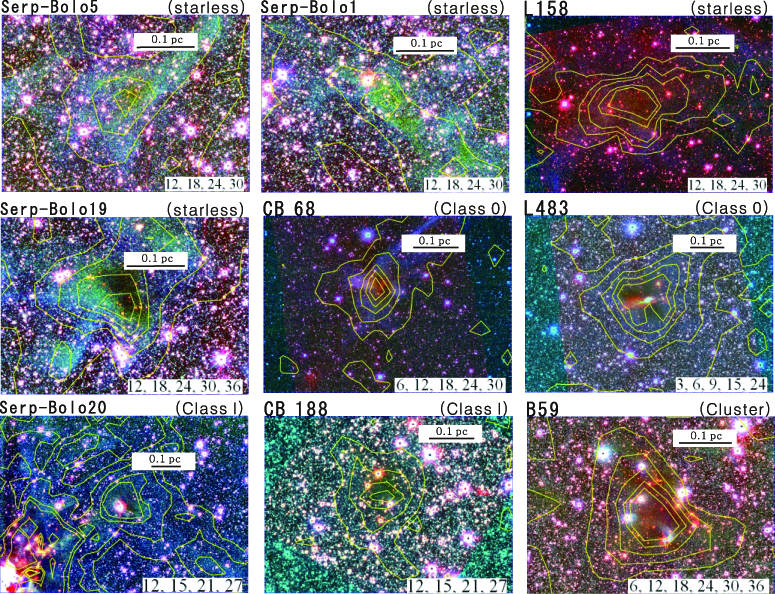}
	\label{fig1}
	\caption{The extinction maps of the nine cores. The pseudocolor images display our CFHT H and Ks band and Spitzer's IRAC1 band data in blue, green, and red respectively. The contours represent the extinction Av and the contour levels are shown in the corner of each image. The black lines represent a the length scale of 0.1 pc. }
\end{figure*}

\begin{figure*}
	\centering
	\includegraphics[width=\textwidth]{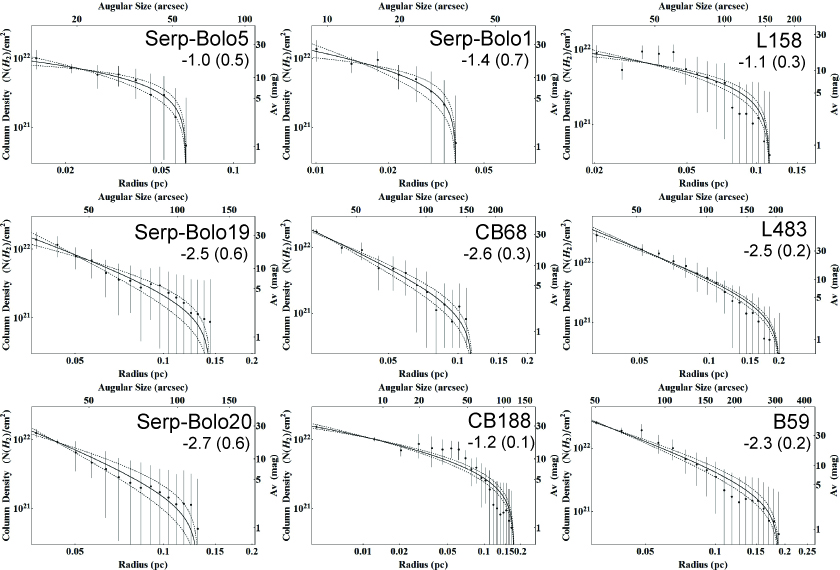}
	\label{fig2}
	\caption{The density profiles of the nine cores.  The figures are the density profile of each core overlaid with the fitting results for the power-law density profile model in black solid lines.  The numbers indicate the best-fit power-law index and its associated uncertainty. The dashed lines represent the uncertainties of best-fit profiles.  The black dots show the density profile with the background extinction subtracted. The error bar is the standard deviation of annularly averaged extinction.  }
\end{figure*}

\begin{figure*}
	\centering
	\includegraphics[height=50mm]{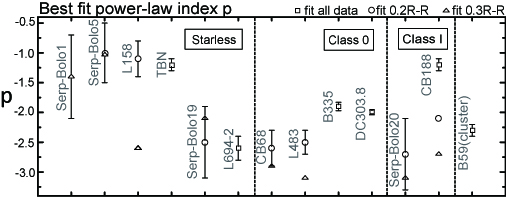}
	\label{fig3}
	\caption{The best fit power-law index of the power-law density model. 
The black squares are the best-fit results, the black circles and the black triangles are obtained with fitting ranges $\sim 0.2R-R$ and $\sim 0.3R-R$, respectively.}
\end{figure*}


\begin{thebibliography}{}
\bibitem[Alves et al.(2001)]{2001Natur.409..159A} Alves, J.~F., Lada, 
C.~J., \& Lada, E.~A.\ 2001, \nat, 409, 159  
\bibitem[Andre et al.(2000)]{2000prpl.conf...59A} Andre, P., Ward-Thompson, 
D., \& Barsony, M.\ 2000, Protostars and Planets IV, 59 


\bibitem[Bertin 
\& Arnouts(1996)]{1996A&AS..117..393B} Bertin, E., \& Arnouts, S.\ 1996, \aaps, 117, 393 
\bibitem[Bohlin et al.(1978)]{1978ApJ...224..132B} Bohlin, R.~C., Savage, 
B.~D., \& Drake, J.~F.\ 1978, \apj, 224, 132 
 \bibitem[Bonnor(1956)]{1956MNRAS.116..351B} Bonnor, W.~B.\ 1956, \mnras, 
116, 351 

\bibitem[Chapman et al.(2009)]{2009ApJ...690..496C} Chapman, N.~L., Mundy, 
L.~G., Lai, S.-P., \& Evans, N.~J.\ 2009, \apj, 690, 496 

 \bibitem[Ebert(1955)]{1955ZA.....37..217E} Ebert, R.\ 1955, Zeitschrift fur 
Astrophysik, 37, 217 
\bibitem[Enoch et al.(2007)]{2007ApJ...666..982E} Enoch, M.~L., Glenn, J., 
Evans, N.~J., II, Sargent, A.~I., Young, K.~E., 
\& Huard, T.~L.\ 2007, \apj, 666, 982 
\bibitem[Evans et al.(2003)]{2003PASP..115..965E} Evans, N.~J., et al.\ 
2003, \pasp, 115, 965

\bibitem[Fatuzzo et al.(2004)]{2004ApJ...615..813F} Fatuzzo, M., Adams, 
F.~C., \& Myers, P.~C.\ 2004, \apj, 615, 813 


\bibitem[Harvey et al.(2001)]{2001ApJ...563..903H} Harvey, D.~W.~A., Wilner, D.~J., Lada, C.~J., Myers, P.~C., Alves, J.~F., \& Chen, H.\ 2001, \apj, 563, 903 
\bibitem[Harvey et al.(2003)]{2003ApJ...596..383H} Harvey, D.~W.~A., Wilner, D.~J., Myers, P.~C., \& Tafalla, M.\ 2003, \apj, 596, 383 
\bibitem[Harvey et al.(2003)]{2003ApJ...598.1112H} Harvey, D.~W.~A., Wilner, D.~J., Lada, C.~J., Myers, P.~C., \& Alves, J.~F.\ 2003, \apj, 598, 1112 
\bibitem[Harvey et al.(2007)]{2007ApJ...663.1149H} Harvey, P., Mer{\'{\i}}n, B., Huard, T.~L., Rebull, L.~M., Chapman, N., Evans, N.~J., II, \& Myers, P.~C.\ 2007, \apj, 663, 1149


\bibitem[Indebetouw et al.(2005)]{2005ApJ...619..931I} Indebetouw, R., et al.\ 2005, \apj, 619, 931 
\bibitem[Kainulainen et al.(2007)]{2007A&A...463.1029K} Kainulainen, J., Lehtinen, K., V{\"a}is{\"a}nen, P., Bronfman, L., \& Knude, J.\ 2007, \aap, 463, 1029 
\bibitem[Kandori et al.(2005)]{2005AJ....130.2166K} Kandori, R., et al.\ 2005, \aj, 130, 2166 
\bibitem[Kauffmann et al.(2008)]{2008A&A...487..993K} Kauffmann, J., Bertoldi, F., Bourke, T.~L., Evans, N.~J., II, \& Lee, C.~W.\ 2008, \aap, 487, 993 


\bibitem[Lada et al.(1994)]{1994ApJ...429..694L} Lada, C.~J., Lada, E.~A., Clemens, D.~P., \& Bally, J.\ 1994, \apj, 429, 694 
\bibitem[Lee et al.(2004)]{2004ApJS..153..523L} Lee, C.~W., Myers, P.~C., \& Plume, R.\ 2004, \apjs, 153, 523 
\bibitem[Lee et al.(2007)]{2007ApJ...660.1326L} Lee, S.~H., Park, Y.-S., Sohn, J., Lee, C.~W., \& Lee, H.~M.\ 2007, \apj, 660, 1326 


\bibitem[Press(1993)]{1993nrc..book.....P} Press, W.~H.\ 1993, Numerical 
recipes in C : the art of scientific computing (Cambridge: Cambridge Univ. Press)


\bibitem[Shirley et al.(2000)]{2000ApJS..131..249S} Shirley, Y.~L., Evans, N.~J., II, Rawlings, J.~M.~C., \& Gregersen, E.~M.\ 2000, \apjs, 131, 249 
\bibitem[Shu(1977)]{1977ApJ...214..488S} Shu, F.~H.\ 1977, \apj, 214, 488 
\bibitem[Stetson(1987)]{1987PASP...99..191S} Stetson, P.~B.\ 1987, \pasp, 99, 191 
\bibitem[Tafalla et 
al.(2000)]{2000A&A...359..967T} Tafalla, M., Myers, P.~C., Mardones, D., \& Bachiller, R.\ 2000, \aap, 359, 967 
\bibitem[Tassis 
\& Mouschovias(2007)]{2007ApJ...660..388T} Tassis, K., \& Mouschovias, T.~C.\ 2007, \apj, 660, 388 


\bibitem[Vall{\'e}e \& Fiege(2007)]{2007AJ....134..628V} Vall{\'e}e, J.~P., \& Fiege, J.~D.\ 2007, \aj, 134, 628 
\bibitem[Vorobyov \& Basu(2005)]{2005MNRAS.360..675V} Vorobyov, E.~I., \& Basu, S.\ 2005, \mnras, 360, 675 

\bibitem[Wallin et al.(2007)]{2007ApJ...666.1296W} Wallin, J.~F., Dixon, 
D.~S., \& Page, G.~L.\ 2007, \apj, 666, 1296 
\bibitem[Weingartner \& Draine(2001)]{2001ApJ...548..296W} Weingartner, J.~C., \& Draine, B.~T.\ 2001, \apj, 548, 296 
\end{thebibliography}
\end{document}